\documentclass{article}

\usepackage{arxiv}

\usepackage[utf8]{inputenc} 
\usepackage[T1]{fontenc}    
\usepackage{hyperref}       
\usepackage{url}            
\usepackage{booktabs}       
\usepackage{amsfonts}       
\usepackage{nicefrac}       
\usepackage{microtype}      
\usepackage{lipsum}
\usepackage{graphicx}
\graphicspath{ {./} }
\usepackage{amsmath}
\usepackage{braket}
\usepackage{hyperref}
\usepackage{amssymb}
\usepackage{subcaption}
\usepackage{amsthm}
\usepackage{mdframed}
\usepackage{tikz}
\usetikzlibrary{positioning, shapes, fit, arrows.meta, calc, decorations.pathreplacing}
\usepackage{mdframed}
\usepackage{comment}
\usepackage{xeCJK}
\usepackage{enumitem}

\title{Non-Markovian effects on informational steady states}

\author{
Jacob Werner \\
  Department of Physics, The University of Tokyo\\
  5-1-5 Kashiwanoha, Kashiwa-shi, Chiba 277-8574, Japan\\
  \texttt{jacob.werner29@gmail.com}
}

\begin{document}
\maketitle
\begin{abstract}
Informational steady states (ISSs) arise when the rate at which information is gained from continuous measurements is exactly balanced by the rate at which information from earlier measurements is lost to the environment \cite{PRXQuantum.3.010303}. While ISSs have been studied in Markovian settings, relatively little is known about how memory effects influence their formation and steady-state properties. In this work, we investigate non-Markovian effects on ISSs within the framework of continuously monitored collision models. We model the environment as a sequence of two-qubit ancillas and find that, for this specific model, the steady-state information gain per measurement is negatively correlated with the degree of non-Markovianity. Numerical results further characterize how the system--ancilla and ancilla--ancilla interaction parameters influence both the degree of non-Markovianity and the steady-state information gain.

\end{abstract}


\section{Introduction}
Throughout this paper, we will model the environment with collision models \cite{Ciccarello_2022}, also known as repeated interaction schemes. Rather than dealing with a complicated environment with many degrees of freedom, the environment is modeled as a series of ancillas that sequentially interact with the system, each possessing only a few degrees of freedom. Despite this simplification, collision models capture many of the key features of open quantum system dynamics. Additionally, their simple structure makes it straightforward to introduce and control non-Markovianity, i.e., information backflow. 

Furthermore, by allowing measurements, collision models provide a natural setting to study the concept of informational steady states (ISSs), which was introduced in Ref.~\cite{PRXQuantum.3.010303}. The idea is that as measurements are made, information is gained, but information from earlier measurements is also lost due to interactions with the environment. In an ISS, the gain and loss are nonzero and balance each other out. 

The concept of ISSs was introduced relatively recently in 2022 \cite{PRXQuantum.3.010303}. To date, relatively little work has been devoted to understanding their properties, and a drawback of all of the current research on ISSs has been that there is no mention of information backflow. However, many realistic environments indeed do possess memory, and hence understanding these effects is important both fundamentally and for future quantum technologies.  Understanding more deeply the information dynamics of a measurement record can also have applications in feedback-control protocols, particularly as experimental control of quantum systems improves.

Additionally, information is increasingly being viewed as a physical quantity, with potential influences on entropy production, work extraction, irreversibility, etc. For example, ISSs were originally discussed in the context of thermodynamics, with a focus on conditional entropy production and irreversibility. It was shown that the steady-state information obtained from a measurement record quantifies the decrease in entropy production for a monitored process \cite{PRXQuantum.3.010303}. Therefore, understanding how these informational quantities are affected by non-Markovianity can be relevant in understanding the energetics and thermodynamics of more realistic continuously monitored systems.

The paper is structured as follows. Sec.~\ref{sec:coll} reviews the general framework of collision models and how non-Markovianity can be introduced through ancilla--ancilla interactions. Sec.~\ref{sec:gain} introduces several quantities relevant to informational steady states. Sec.~\ref{sec:nonm} discusses non-Markovian effects on ISSs, presenting numerical results. Finally, we summarize our findings and conclude in Sec.~\ref{sec:conc}.


\section{Collision models}
\label{sec:coll}
\subsection{General framework}
Collision models (CMs) are widespread in open quantum systems research, because despite their simplicity, they allow for the characterization of a wide range of phenomena. In a CM, the system-environment interaction occurs in collisions, and each collision designates a discrete time step \cite{Ciccarello_2022}. That is, when we say an event occurs at t = 1, we mean that the event occurs during or directly after the first collision, depending on context.

\begin{figure}
\centering
\includegraphics[scale=0.6]{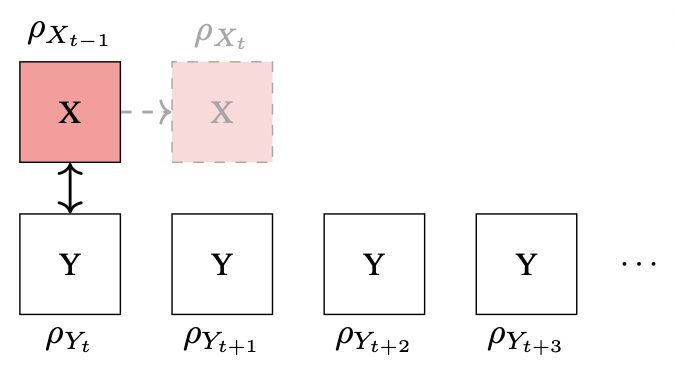}
\caption{Markovian CM schematic}
\label{fig:schematic1}
\end{figure}

In a canonical Markovian CM, each incoming ancilla ($Y$) is not correlated to the system ($X$) or to other ancillas. A schematic for this process is shown in Fig.~\ref{fig:schematic1}. The system at time $t$, i.e., after the $t$th collision, is given by

\begin{equation}
\rho_{X_t} = \mathcal{E}_t[\rho_{X_{t-1}}] = \operatorname{Tr}_{Y_t}\big[ U_{X_{t-1}Y_t}  (\rho_{X_{t-1}} \otimes \rho_{Y_t} ) U_{X_{t-1}Y_t}^\dagger \big],
\label{eq:markcm}
\end{equation}

\noindent where $\rho_{Y_t}$ is the $t$th ancilla, and $U_{X_{t-1}Y_t}$ is the unitary governing the $t$th system--ancilla ($XY$) interaction. It is apparent that CMs are just repeated CPTP maps.

Throughout this paper, we will restrict ourselves to time-independent unitaries and incoming ancillas, i.e., $U_{X_{t-1}Y_t}=U_{XY}$, $\rho_{Y_t}=\rho_Y$, and $\mathcal{E}_t=\mathcal{E}$. Additionally, all interactions will be partial swaps, parameterized by $\theta$ and expressed as

\begin{equation}
U(\theta) = e^{i \theta S}=\cos(\theta)\,\mathbb{I} + i \sin(\theta)\,S,
\end{equation}

\noindent where $\mathbb{I}$ denotes the $4 \times 4$ identity operator, and $S$ is the SWAP operator. In the computational basis ($\sigma_z$), this is given by

\begin{equation}
S =
\begin{pmatrix}
1 & 0 & 0 & 0 \\
0 & 0 & 1 & 0 \\
0 & 1 & 0 & 0 \\
0 & 0 & 0 & 1
\end{pmatrix}.
\end{equation}

\noindent
For clarity, the partial swap on a couple states is given below:
\begin{align}
U(\theta) \, \lvert 00 \rangle &= e^{i\theta} \lvert 00 \rangle, \\
U(\theta) \, \lvert 10 \rangle &= \cos(\theta)\, \lvert 10 \rangle + i \sin(\theta)\, \lvert 01 \rangle.
\end{align}

\noindent When $\theta=0$, there is no interaction, and when $\theta=\pi/2$, the interaction is a full swap. 

The partial swap is used extensively in the literature on CMs for several reasons. Firstly, after two systems interact, they each retain a clear imprint of the system they just interacted with. Therefore, if mechanisms for non-Markovianity are introduced, information backflow can often be inferred simply from the state of the system, for example, if it becomes more similar to past system states. Secondly, partial swaps have a clear physical interpretation, being generated from the Heisenberg exchange interaction:

\begin{equation}
H_{XY}
= -J\Big( \boldsymbol{\sigma}_X \cdot \boldsymbol{\sigma}_Y \Big)
= -J\Big(\sigma_x \otimes \sigma_x
+ \sigma_y \otimes \sigma_y
+ \sigma_z \otimes \sigma_z
\Big),
\label{eq:heisenberg}
\end{equation}

\noindent where $J$ is a constant determining the strength of the interaction. If each collision lasts for $\Delta t$, then the corresponding value of $\theta$ is given by $\theta=2J\Delta t$. 

Additionally, take the system and ancilla local Hamiltonians to be $H_X = \hbar \omega_X \sigma_z$ and $H_Y = \hbar \omega_Y \sigma_z$, respectively. If they are resonant, i.e., $\omega_X=\omega_Y$, then $[H_{XY},H_X+H_Y]=0$, and in the interaction picture, we only need to consider $H_{XY}$. We will therefore assume resonance throughout this investigation and work in the interaction picture. Therefore, the total Hamiltonian can be simplified as follows:

\begin{equation}
H(t_0)
= H_X \;+\; \sum_{t=1}^{\infty} H_{Y_t}
\;+\; \sum_{t=1}^{\infty} \chi_t(t_0)\, H_{X Y_t}
\longrightarrow
H_I(t_0)
= \sum_{t=1}^{\infty} \chi_t(t_0)\, H_{X Y_t},
\end{equation}

\noindent where $\chi_t(t_0)$ is $1$ if the $t$th ancilla is interacting and $0$ otherwise. 

\subsection{Non-Markovian collision models}
If we also allow ancilla--ancilla ($YY$) interactions, there is now a clear mechanism for information backflow \cite{Ciccarello_2022}. For example, if there is a nearest-neighbor $YY$ interaction after the $XY$ interaction, the state at time $t$, $\rho_{X_t}$, is given by

\begin{align}
\rho_{X_t} &= \mathcal{E}_t\big[\rho_{X_{t-1}}\big]  \nonumber \\
&= \operatorname{Tr}_{Y_t,Y_{t+1}}\big[ U_{YY}U_{XY} \left( \rho_{X_{t-1}Y_{t}} \otimes \rho_{Y_{t+1}}\right) U_{XY}^\dagger U_{YY}^\dagger\big],
\label{eq:nonm}
\end{align}

\noindent where $U_{XY}$ and $U_{YY}$ are parameterized by $\theta_{XY}$ and $\theta_{YY}$, respectively. Before the $XY$ interaction occurs, the system is now in general already correlated to the ancilla. The joint system--ancilla state at time $t$, $\rho_{X_tY_{t+1}}$, is given by the same expression above, but only taking the trace over $Y_t$. This is the joint system--ancilla state, where the ancilla is the incoming ancilla. To obtain the joint system--ancilla state where the ancilla is the outgoing ancilla, trace over $Y_{t+1}$. 

Due to the correlations, the map is now, in general, non-CPTP and time-dependent. Also, the new Hamiltonian is given by

\begin{equation}
H_I(t_0)
= \sum_{t=1}^{\infty} \chi_t(t_0)\, H_{X Y_t}
+ \sum_{t=1}^{\infty} \eta_t(t_0)\, H_{Y_t Y_{t+1}},
\end{equation}

\noindent where $\eta_t(t_0)$ is $1$ during the corresponding $YY$ interaction and $0$ otherwise. Note that if we only consider partial swaps, we still have $[H_I,H_X \;+\; \sum_{t=1}^{\infty} H_{Y_t}]=0$. 

\subsection{Quantifying non-Markovianity}
While the definition of classical Markovianity is well-defined through conditional probabilities, there is no clear consensus on the definition for quantum processes. However, all definitions aim to capture some concept of memory effects or information backflow from the environment. It is often said that in a Markovian process,
the system continually loses information to the environment, while there is an exchange of
information in a non-Markovian process \cite{Breuer_2016}. The information that goes to the environment
and later comes back is stored in correlations or changes to the environmental states.

Throughout this paper, we will use the Breuer-Laine-Piilo (BLP) measure of non-Markovianity, which proves very useful for discrete-time dynamical processes. Introduced in Ref.~\cite{Breuer_2009}, the measure is based on the idea that Markovian processes should reduce the distinguishability between any two states. For two states $\rho_1$ and $\rho_2$, their trace distance is used to quantify their distinguishability, given by

\begin{equation}
D\left( \rho_1, \rho_2 \right) = \frac{1}{2} \left\| \rho_1 - \rho_2 \right\|_1.
\end{equation}

The BLP measure $\mathcal{N}_{\mathrm{BLP}}$ is then given by summing over all increases in the trace distance:

\begin{equation}
\mathcal{N}_{\mathrm{BLP}}
= \max_{\rho_{1,0}, \rho_{2,0}}
\sum_{t:\, \Delta D > 0}
\Big[
D(\rho_{1,t}, \rho_{2,t})
- D(\rho_{1,t-1}, \rho_{2,t-1})
\Big].
\label{eq:discrete}
\end{equation}

For ease of computation, rather than taking a maximization over all initial state pairs, we will simply use the $\sigma_x$ eigenstates to determine $\mathcal{N}_{\mathrm{BLP}}$. For the qualitative conclusions of this paper, the difference is negligible. 

Note that Markovian CMs are a series of CPTP maps, so that the trace distance between two states continually decreases over time and $\mathcal{N}_{\mathrm{BLP}}=0$ \cite{Lindblad1975}. However, the simplicity of CMs makes them a perfect foundation on which to introduce various non-Markovian mechanisms.

\subsection{Continuously monitored collision models}
This paper concerns itself with informational steady states (ISSs), which only occur in continuously monitored collision models (CM$^2$), where measurements are made at every time step \cite{PRXQuantum.3.010303}. We will refer to a state conditioned on a specific measurement record as a conditional state.

In a CM$^2$, the measurements are performed on an ancilla after all of its interactions, indirectly giving us information about the system. First, consider the Markovian case, with no $YY$ interactions. At time $t$, the conditional joint state of the system and the outgoing ancilla is given by

\begin{equation}
\rho_{X_{t}Y_t|\zeta_{t-1}} = U_{XY} (\rho_{X_{t-1}|\zeta_{t-1}} \otimes \rho_{Y_t}) U_{XY}^\dagger.
\end{equation}

\noindent Then, a POVM is performed on the ancilla, described by operators satisfying $\sum_z M_z^\dagger M_z = \mathbb{I}$. Outcome $z_t$ occurs with probability $P(z_t)$ given by

\begin{equation}
P(z_t) = \operatorname{Tr} \big[ M_{z_t} \rho_{Y_t}' M_{z_t}^\dagger \big].
\end{equation}

\noindent The state $\rho_{Y_t}'$ is the ancilla after the interaction, given by tracing over the system, i.e., $\rho_{Y_t}'=\operatorname{Tr}_{X}\big[\rho_{X_tY_t|\zeta_{t-1}}\big]$. The probability of a joint measurement record $\zeta_t = \{z_1, \ldots, z_t\}$ is given by

\begin{equation}
P(\zeta_t) = \operatorname{Tr}_{X Y_1 \ldots Y_t} \big[ M_{z_t} \cdots M_{z_1} \rho_{X_t Y_1 \ldots Y_t} M_{z_1}^\dagger \cdots M_{z_t}^\dagger \big],
\label{eq:sys}
\end{equation}

\noindent where $\rho_{X_t Y_1 \ldots Y_t}$ is the unconditional state. This is given by

\begin{equation}
\rho_{X_t Y_1 \ldots Y_t} = \left( \prod_{i=1}^t U_{XY_i} \right)
\left( \rho_{X_0} \otimes \bigotimes_{k=1}^t \rho_{Y_k} \right)
\left( \prod_{j=1}^t U_{XY_j} \right)^\dagger.
\end{equation}

\noindent To be convinced of Eq.~\eqref{eq:sys}, remember that the measurement operators only act on an ancilla \textit{after} all of its interactions. Therefore, a given $M_z$ commutes with all future unitaries. Also, if an ancilla is done interacting, tracing it out at a later time will give the same system state. Therefore, the conditional system state is given by

\begin{equation}
\rho_{X_t | \zeta_t} = \frac{1}{P(\zeta_t)} \, \operatorname{Tr}_{Y_1 \ldots Y_t} \Bigg[ 
\left( \prod_{i=1}^t M_{z_i} \right) 
\rho_{X_t Y_1 \ldots Y_t} 
\left( \prod_{j=1}^t M_{z_j} \right)^\dagger 
\Bigg].
\end{equation}

\noindent Also, due to the no-signaling theorem \cite{nielsen00}, the average of the conditional states must give the unconditional state:

\begin{equation}
\sum_{\zeta_t} P(\zeta_t) \rho_{X_t | \zeta_t} = \rho_{X_t}.
\label{eq:signalling}
\end{equation}

\subsection{Two-qubit ancilla collision model}
\begin{figure}
\centering
\includegraphics[scale=0.6]{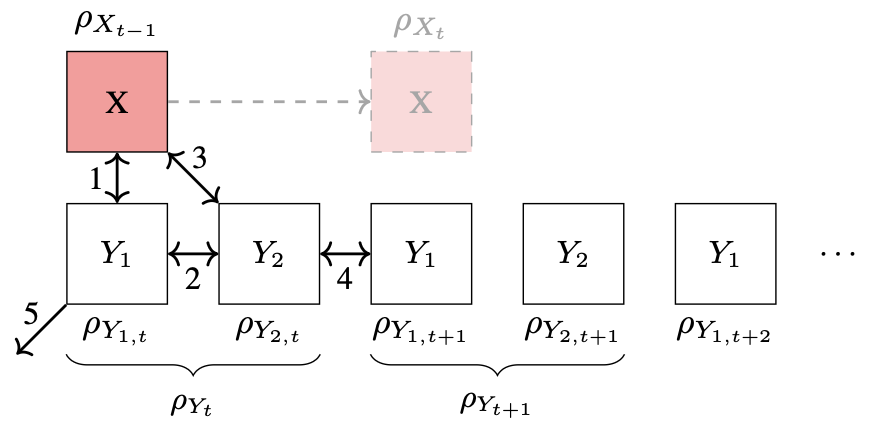}
\caption{Two-qubit ancilla CM schematic}
\label{fig:schematic2}
\end{figure}

Throughout this paper, we will consider two-qubit ancillas to ensure an informational steady state is approached where the steady-state gain and loss of information are nonzero. A schematic is shown in Fig.~\ref{fig:schematic2}.

The ancillas are given by $\rho_{Y_t}=\rho_{Y_{1,t}}\otimes\rho_{Y_{2,t}}$. Each time step consists of five operations, the final being a measurement on $\rho_{Y_{1,t}}$. In this paper, each $\rho_{Y_{1,t}}$ will be initialized to a thermal state which is diagonal in the $\sigma_z$ basis. We will parameterize them by their excited state populations, i.e., $\bra{0}\rho_{\mathrm{th}}(a)\ket{0}=a$ and $\bra{1}\rho_{\mathrm{th}}(a)\ket{1}=1-a$. Here we will set $\rho_{Y_{1,t}}=\rho_{\mathrm{th}}(0.3)$. Each $\rho_{Y_{2,t}}$ will be initialized to $\ket{+}\bra{+}$, where $\ket{+}$ is the $\sigma_x$ positive eigenstate. The first and second $XY$ interactions (operations 1 and 3) will be parameterized by $\theta_{XY_1}$ and $\theta_{XY_2}$, respectively. Note that operation 2 occurs within one ancilla and hence is technically not an ancilla--ancilla interaction. However, we will still label this a $YY$ interaction. We will parameterize operations 2 and 4 by $\theta_{Y_1Y_2}$ and $\theta_{Y_2Y_1}$, respectively. Note that measurements are made only on $\rho_{Y_{1,t}}$, with $M_0=\ket{0}\bra{0}$ and $M_1=\ket{1}\bra{1}$.

\section{Gain and Loss}
\label{sec:gain}
As mentioned previously, an ISS occurs if the gain of information from measurements and the loss of information from interactions with the environment are nonzero and equal in the long-time limit. The conditional evolution includes measurements and is stochastic, whereas the unconditional evolution is deterministic. Shown in Eq.~\eqref{eq:signalling}, the average of the conditional states gives the unconditional state, but the same is not necessarily true for the von Neumann entropy. The unconditional entropy and the conditional entropy are given, respectively, by

\begin{align}
S(X_t) &= - \operatorname{Tr} \, \rho_{X_t} \ln \rho_{X_t},\\
S(X_t | \zeta_t) &= \sum_{\zeta_t} P(\zeta_t) S(\rho_{X_t | \zeta_t}).
\end{align}

The difference between the two quantities takes the form of a Holevo information \cite{nielsen00}, which we denote as follows:

\begin{equation}
I(X_t : \zeta_t) := S(X_t) - S(X_t | \zeta_t).
\end{equation}

\noindent Therefore, the Holevo information at time $t$ is the information that the measurement record up to time $t$ provides about the state at time $t$. Although the Holevo information is always non-negative, the Holevo rate, i.e., the change in the Holevo information per time step, given by $\Delta I_t := I(X_t : \zeta_t) - I(X_{t-1} : \zeta_{t-1})$, can take any sign. We can split this change up into a gain and a loss term, i.e., $\Delta I_t=G_t-L_t$. 

The most recent measurement outcome $z_{t}$ provides more information about the current state. Mathematically, this gain is given by

\begin{align}
G_t &= I(X_t : \zeta_t) - I(X_t : \zeta_{t-1})\\
&= S(X_t | \zeta_{t-1}) - S(X_t | \zeta_t)\\
&=\sum_{\zeta_t}P(\zeta_t)D(\rho_{X_t|\zeta_t}\|\rho_{X_t|\zeta_{t-1}}).
\end{align}

\noindent The last line shows that the gain is non-negative for both CPTP and non-CPTP maps. To better understand the gain, consider the classical quantum mutual information (CQMI) between systems $A$ and $B$, given by

\begin{equation}
J(\rho_{AB})=S(\rho_A)-S_M(\rho_{A}|\rho_B),
\label{eq:cqmi}
\end{equation}

\noindent where $S_M(\rho_{A}|\rho_B)=\sum_i P(i)S(\rho_{A,i})$ is a quantum generalization of the classical conditional entropy \cite{nielsen00}. Here, $\rho_{A,i}$ is $A$'s state conditioned on a measurement $i$ on $B$. This definition depends on the POVM $\{M_i\}$ used to measure $B$. Since this relies on a classical measurement outcome, $J(\rho_{AB})$ is often said to quantify the classical correlations between two quantum systems. The gain $G_t$ can be rewritten as the average over the $\zeta_{t-1}$ distribution of the final CQMI of the system and $Y_1$. In other words, the gain quantifies the classical correlations formed between the system and $Y_1$.

Regarding the loss, we expect the measurement record up to time $t-1$, $\zeta_{t-1}$, to contain less information about the system at time $t$ than at time $t-1$. This loss is given by

\begin{align}
L_t &= I(X_{t-1} : \zeta_{t-1}) - I(X_t : \zeta_{t-1})\nonumber \\
&=\sum_{\zeta_{t-1}} P(\zeta_{t-1}) \left[ 
D(\rho_{X_{t-1} | \zeta_{t-1}} \| \rho_{X_{t-1}}) 
- D(\rho_{X_t | \zeta_{t-1}} \| \rho_{X_t}) 
\right].
\label{eq:loss}
\end{align}

\noindent Note that $\rho_{X_{t} | \zeta_{t-1}}=\mathcal{E}_t[\rho_{X_{t-1} | \zeta_{t-1}}]$ and $\rho_{X_t}=\mathcal{E}_t[\rho_{X_{t-1}}]$. For CPTP maps, the data processing inequality guarantees that the relative entropy between two states cannot increase \cite{Lindblad1975}. Therefore, Eq.~\eqref{eq:loss} is a sum of non-negative terms for CPTP maps. 

Note that for a qubit system, $I(X_t:\zeta_t)$ is upper bounded by $\ln2\approx0.693$ and lower bounded by $0$. Consequently, unless a limit cycle is formed, $\Delta I_t=0$ must hold in the long-time limit, i.e., $G_{\infty}=L_{\infty}$. It follows that any argument that we make for the value of $G_{\infty}$ also applies to  $L_{\infty}$. Therefore, to reference the value that an ISS settles at, we will simply use $G_{\infty}$ from now on. Note that if the same map is repeated, there should not be a limit cycle, even with $YY$ interactions. 

An ISS is defined such that $\Delta I_{\infty}=0$, and $G_{\infty}=L_{\infty}\neq0$. In other words, information is continuously gained and continuously lost such that the current measurement record always provides the same amount of information about the current system. The situation where $G_{\infty}=L_{\infty}=0$ can be referred to as informational equilibrium (IE).

\begin{figure}
\centering
\includegraphics[scale=0.65]{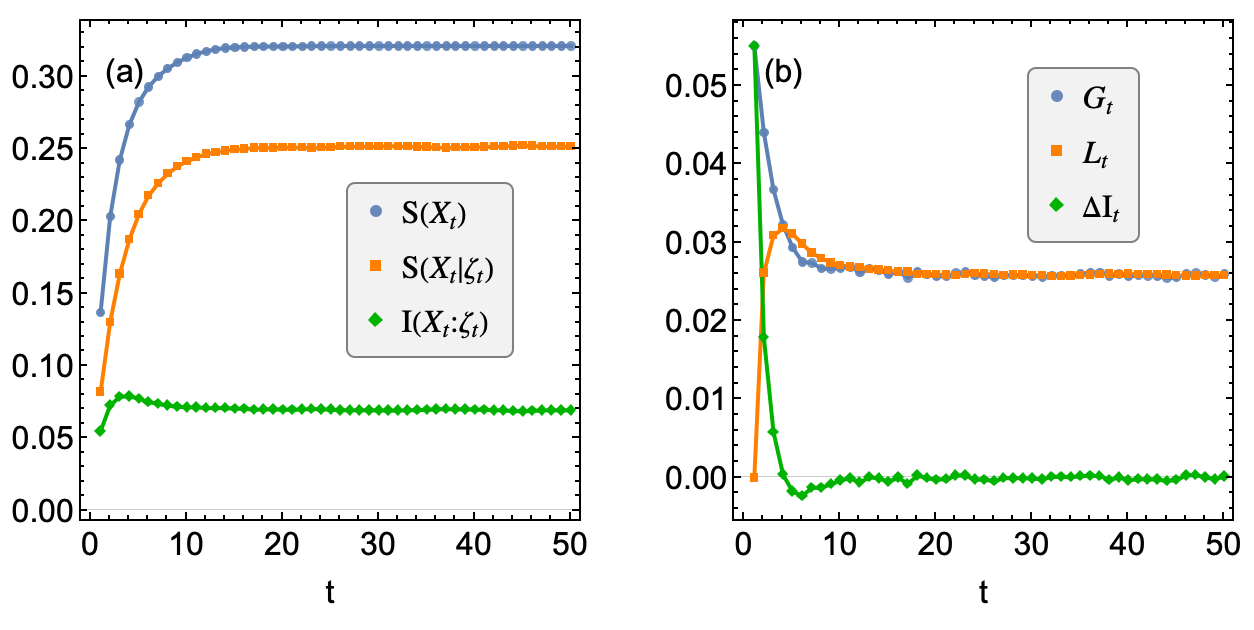}
\caption{Key conditional quantities are shown, averaged over 10,000 trajectories, with $\rho_{X_0}=\ket{+}\bra{+}$, $\rho_Y=\rho_{\mathrm{th}}(0.3)\otimes\ket{+}\bra{+}$, $\theta_{XY_1}=0.3$, $\theta_{XY_2}=0.5$, and $\theta_{Y_1Y_2}=\theta_{Y_2Y_1}=0$. Results are reproduced from Ref.~\cite{PRXQuantum.3.010303}.}
\label{fig:model2fig}
\end{figure}

For our model, the key quantities are plotted in Fig.~\ref{fig:model2fig}, reproduced from Ref.~\cite{PRXQuantum.3.010303}. Since we use a two-qubit ancilla, the two qubits will compete to determine the steady state, and the unconditional process embodies a NESS with continuous entropy production \cite{PRXQuantum.3.010303}. The conditional process embodies an ISS, and from the figure, we observe that this is reached relatively quickly.

\section{Non-Markovian effects on ISSs}
\label{sec:nonm}


In this section, we present numerical results showing the effects of varying the $XY$ and $YY$ interaction parameters on $\mathcal{N}_{\mathrm{BLP}}$ and $G_{\infty}$. For each parameter setting,  $G_{\infty}$ was obtained by first averaging over the final 30 time steps of each trajectory and then averaging over 20 trajectories, each of length 100 time steps. These values were necessarily kept relatively small due to computational constraints. 

Consider intuitively what we would expect to happen to the gain and loss values as the process becomes non-Markovian. If we introduce a mechanism for information to flow back to the system, we would expect there to be less information available for a measurement to obtain. Therefore, we would likely expect the gain to decrease. If less information is obtained, there should be less available to lose, and hence we also expect the loss to decrease. We quantify this relationship by reporting Spearman's $\rho$ between $G_{\infty}$ and $\mathcal{N}_{\mathrm{BLP}}$ for the non-Markovian processes. Spearman's $\rho$ is a common measure of the strength and direction of the monotonic relationship between two variables, based on the ranks of their values rather than the actual values \cite{Kendall1990}. 






\subsection{Varying $\theta_{XY_1}$ and $\theta_{XY_2}$}
\label{sec:varyingxy}

\begin{figure}
\centering
\includegraphics[scale=0.6]{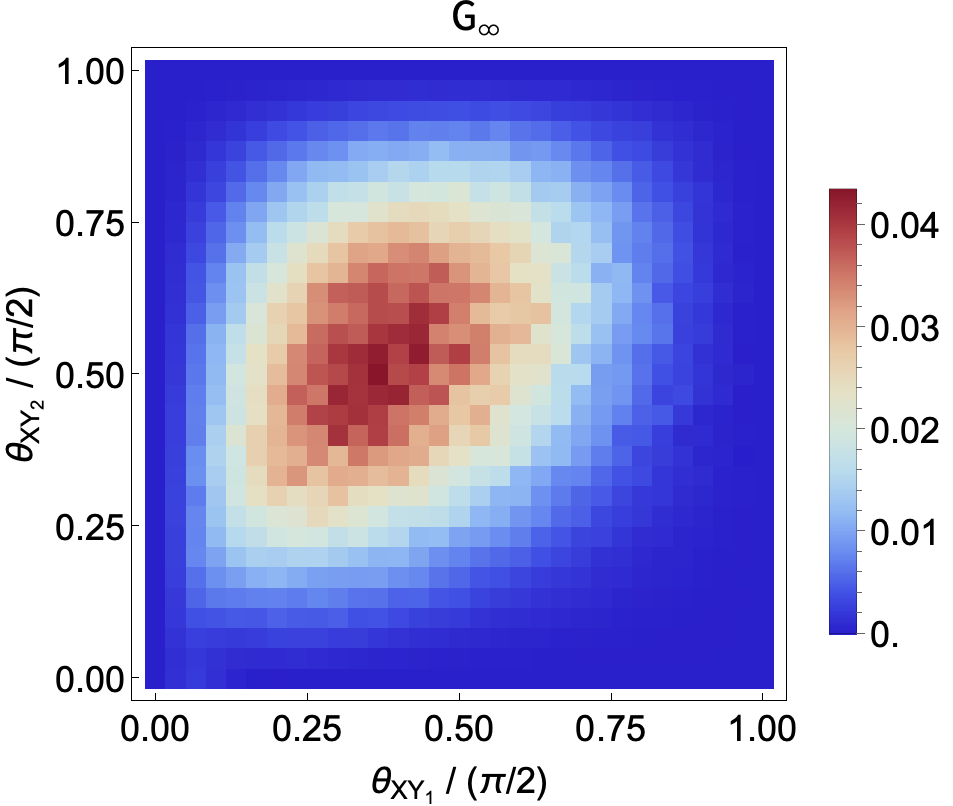}
\caption{$G_{\infty}$ is shown for various values of $\theta_{XY_1}$ and $\theta_{XY_2}$, with $\rho_{X_0}=\ket{+}\bra{+}$ and $\rho_Y=\rho_{\mathrm{th}}(0.3)\otimes\ket{+}\bra{+}$. The $YY$ interaction parameters are set to 0, i.e., the process is Markovian and $\mathcal{N}_{\mathrm{BLP}}=0$ always.}
\label{fig:finals1}
\end{figure}

We first focus on the effects of varying the $XY$ interaction parameters, remaining in the Markovian regime, with $\theta_{Y_1Y_2}=\theta_{Y_2Y_1}=0$. We will provide a qualitative explanation of the $G_{\infty}$ results, shown in Fig.~\ref{fig:finals1}. At the start of a time step, $X$ and $Y_1$ are uncorrelated. If there were no interaction, i.e., $\theta_{XY_1}=0$, they would remain uncorrelated, but a full swap, i.e., $\theta_{XY_1}=\pi/2$, would also keep them uncorrelated. Therefore, we would expect a partial swap with $\theta_{XY_1}\approx\pi/4$ to maximally correlate $X$ and $Y_1$. 

Also, $Y_1$ and $Y_2$ are always uncorrelated before the $XY_2$ interaction. Therefore, increasing $\theta_{XY_2}$ from $0$ to $\pi/2$ should decrease the correlations between $X$ and $Y_1$. Hence, we may expect that $G_{\infty}$ would be maximized for $\theta_{XY_1}\approx\pi/4$ and $\theta_{XY_2}\approx0$. For a given time step and system state, it is true that the CQMI between $X$ and $Y_1$ will likely be maximized for those parameter values. 

However, we have ignored the effect of changing the system state on the CQMI. Recall that $G_{\infty}$ is the average CQMI in the conditional steady state. With $\theta_{XY_1}\approx\pi/4$ and $\theta_{XY_2}\approx0$, the steady state of $X$ will simply be the state $Y_1$, and no correlations will form. Loosely speaking, the greater the difference between $Y_1$ and the steady state of $X$, the greater the steady-state CQMI will be. 

Varying $\theta_{XY_2}$ produces two competing effects. Increasing $\theta_{XY_2}$ from $0$ to $\pi/2$ moves the steady state of $X$ further from $Y_1$ and closer to $Y_2$, which tends to increase the steady state $XY_1$ correlations. At the same time, it causes the subsequent $XY_2$ interaction to erase a larger fraction of the generated correlations, decreasing the steady state $XY_1$ correlations. Varying $\theta_{XY_1}$ also produces two effects. Increasing to $\pi/4$ increases the generated correlations for a given state, and then increasing past $\pi/4$ decreases the correlations. The second effect is that increasing $\theta_{XY_1}$ over the whole range of $0$ to $\pi/2$ moves the steady state closer to $Y_1$. 

Therefore, the maximum of $G_{\infty}$ should occur at $\theta_{XY_2}\approx\pi/4$, assuming the two effects of varying $\theta_{XY_2}$ are similar in strength, and at a value of $\theta_{XY_1}$ less than but relatively close to $\pi/4$. Correspondingly, values above the diagonal $\theta_{XY_1}=\theta_{XY_2}$ of Fig.~\ref{fig:finals1} should be slightly larger compared to those below. This is exactly what is observed.

\subsection{Varying $\theta_{Y_1Y_2}$ and $\theta_{Y_2Y_1}$}
\begin{figure}[t]
\centering
\includegraphics[scale=0.49]{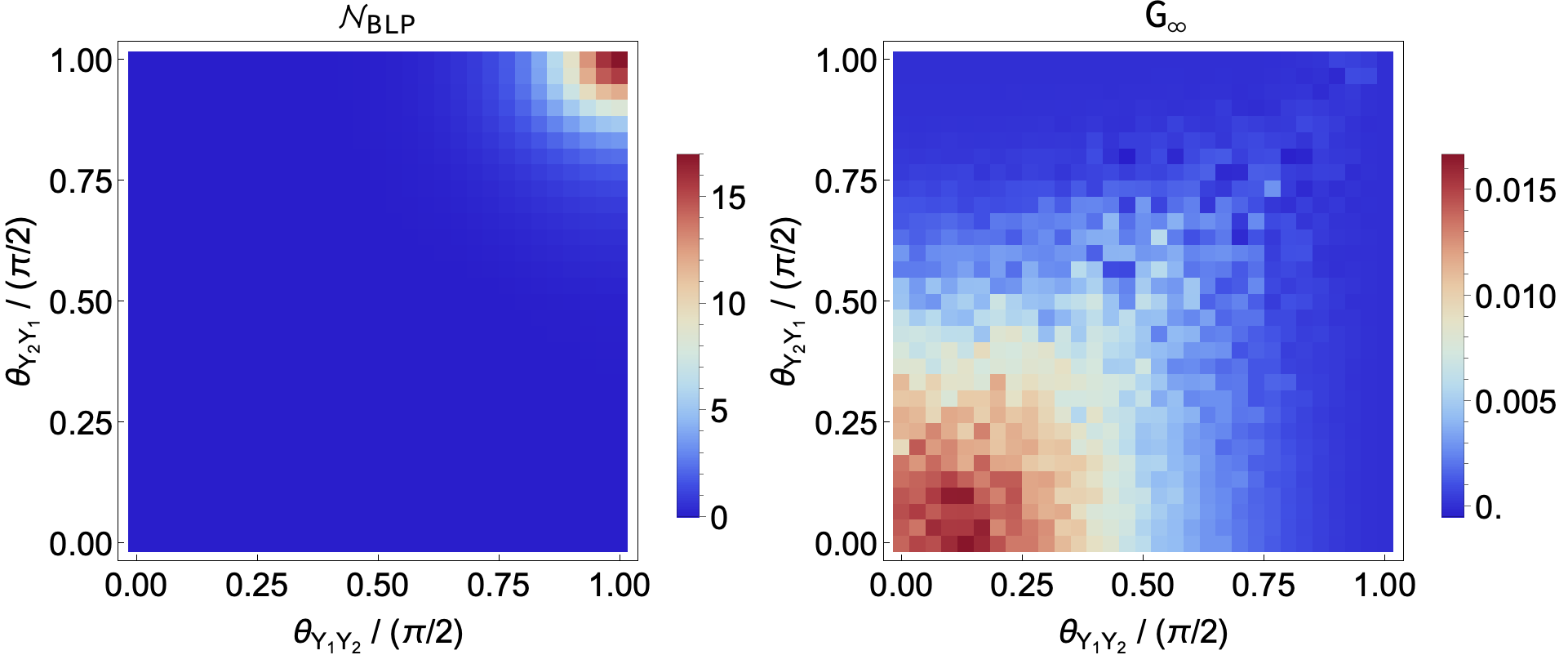}

\vspace{0.3cm}

\includegraphics[scale=0.49]{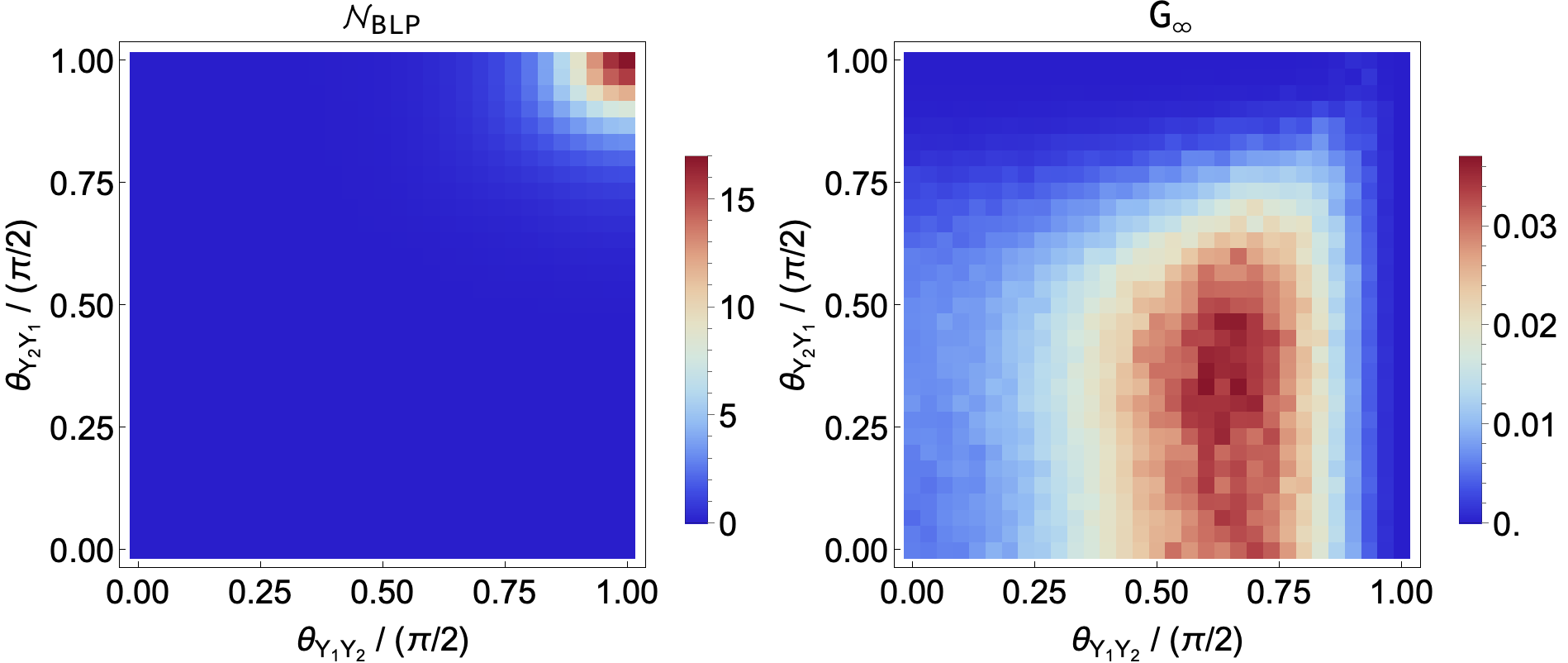}

\caption{$\mathcal{N}_{\mathrm{BLP}}$ and $G_{\infty}$ are shown for various values of $\theta_{Y_1Y_2}$ and $\theta_{Y_2Y_1}$, with $\rho_{X_0}=\ket{+}\bra{+}$ and $\rho_Y=\rho_{\mathrm{th}}(0.3)\otimes\ket{+}\bra{+}$. The upper plots set $\theta_{XY_1}=\theta_{XY_2}=0.2(\pi/2)$, and the lower plots set $\theta_{XY_1}=\theta_{XY_2}=0.8(\pi/2)$. Spearman's $\rho$ for $G_{\infty}$ and $\mathcal{N}_{\mathrm{BLP}}$ is approximately $-0.33$ ($p<10^{-20}$) for $\theta_{XY_1}=\theta_{XY_2}=0.2(\pi/2)$ and $-0.52$ ($p<10^{-20}$) for $\theta_{XY_1}=\theta_{XY_2}=0.8(\pi/2)$.}
\label{fig:finals1011}
\end{figure}

Next, we focus on the effects of varying the $YY$ interaction parameters, keeping $\theta_{XY_1}$ and $\theta_{XY_2}$ fixed and equal to each other. Since the process is in general non-Markovian, we will also discuss how $\mathcal{N}_{\mathrm{BLP}}$ depends on the parameters. The results are shown in Fig.~\ref{fig:finals1011} for both $\theta_{XY_1}=\theta_{XY_2}=0.2(\pi/2)$ (upper plots) and $\theta_{XY_1}=\theta_{XY_2}=0.8(\pi/2)$ (lower plots).

\subsubsection{\boldmath $\mathcal{N}_{\mathrm{BLP}}$}
Focusing on $\mathcal{N}_{\mathrm{BLP}}$, recall that increases in trace distance are due to changes to environmental states and initial correlations with the environment. Considering that each ancilla is composed of two qubits, if $\theta_{Y_1Y_2}>0$ but $\theta_{Y_2Y_1}\approx 0$, then the incoming ancilla will always be unchanged and uncorrelated to the system. Therefore, the trace distance must always decrease at every time step. 

If instead $\theta_{Y_1Y_2}\approx 0$ and $\theta_{Y_2Y_1}>0$, the $Y_2Y_1$ interaction can now change the incoming ancilla and generate correlations. However, if we consider the model as a two-qubit ancilla composed of first $Y_2$, then $Y_1$, we are in the identical situation as before. In this offset model, we are effectively computing the trace distance at times $t=0.5,1.5,2.5,...$, and it must always decrease. Therefore, although possible, it seems unlikely that the trace distance in the original model at times $t=0,1,2,...$ would increase. 

Therefore, if only one of $\theta_{Y_1Y_2}$ and $\theta_{Y_2Y_1}$ is approximately zero, we are constantly blocking the transfer of correlations and information. The left-hand plots of Fig.~\ref{fig:finals1011} confirm that, as intuitively expected, we require large values of both $\theta_{Y_1Y_2}$ and $\theta_{Y_2Y_1}$ to reach large degrees of non-Markovianity.

\subsubsection{\boldmath $G_\infty$}
Firstly, similar to $\theta_{XY_1}$ and $\theta_{XY_2}$, $\theta_{Y_1Y_2}$ and $\theta_{Y_2Y_1}$ have a considerable effect on the steady state of $X$. If $\theta_{Y_1Y_2}$ ($\theta_{Y_2Y_1}$) is large and $\theta_{Y_2Y_1}$ ($\theta_{Y_1Y_2}$) is small, the system will effectively interact with $Y_1$ ($Y_2$) during both the $XY_1$ and $XY_2$ interactions. Therefore, $X$ will tend to a state similar to $Y_1$ ($Y_2$). Also, we have set $\theta_{XY_1}=\theta_{XY_2}$, removing the competition from before and in effect isolating the effects of the $YY$ interaction parameters. 

If $\theta_{Y_1Y_2}$ ($\theta_{Y_2Y_1}$) is large and $\theta_{Y_2Y_1}$ ($\theta_{Y_1Y_2}$) is small, the system tends to an equilibrium with $Y_1$ ($Y_2$), effectively interacting with $Y_1$ ($Y_2$) during both the $XY_1$ and the $XY_2$ interaction. Therefore, the correlations formed with the environment will be minor in the long-time limit, and hence $G_{\infty}$ will be small. This is observed in Fig.~\ref{fig:finals1011}, with both right-hand plots showing values near zero in the top left-hand corner ($\theta_{Y_2Y_1}$ large, $\theta_{Y_1Y_2}$ small) and the bottom right-hand corner ($\theta_{Y_1Y_2}$ large, $\theta_{Y_2Y_1}$ small). 

Also, note that when $\theta_{Y_1Y_2}$ is large, regardless of the value of $\theta_{Y_2Y_1}$, most of the correlations formed during the $XY_1$ interaction are lost. For instance, when $\theta_{Y_1Y_2}=\pi/2$, all correlations will be lost during the $Y_1Y_2$ interaction, and none will be recovered during subsequent interactions. Therefore, the gain will always be zero. 

As mentioned before, we may expect that increasing the degree of non-Markovianity would decrease $G_{\infty}$, as more information flowing back to the system would mean less information is available to be measured. This heuristic does not always hold, as the actual situation is much more complex. It depends on how the degree of non-Markovianity is varied. We should treat the parameters of the model as the cause of changes to $\mathcal{N}_{\mathrm{BLP}}$ and $G_{\infty}$, and not $\mathcal{N}_{\mathrm{BLP}}$ as the cause of changes to $G_{\infty}$. However, here we do indeed see a statistically significant negative correlation between the two quantities.

\subsubsection{\boldmath $G_\infty$ continued: A subtle point}
We now focus on an interesting property, and for simplicity we consider the case of increasing $\theta_{Y_1Y_2}$ when $\theta_{Y_2Y_1}=0$ (bottom rows of the plots of Fig.~\ref{fig:finals1011}). For $\theta_{XY_1}=\theta_{XY_2}=0.2(\pi/2)$, there is a slight increase of $G_{\infty}$ for very small $\theta_{Y_1Y_2}$, but in general, increasing $\theta_{Y_1Y_2}$ leads to a decrease in $G_{\infty}$. However, for $\theta_{XY_1}=\theta_{XY_2}=0.8(\pi/2)$, $G_{\infty}$ peaks at $\theta_{Y_1Y_2}\approx0.6(\pi/2)$. We now qualitatively explain this.


For both settings of $\theta_{XY_1}=\theta_{XY_2}$, the $XY_1$ interaction will increase the $XY_1$ correlations. For $\theta_{Y_1Y_2}\approx0$, there is no $Y_1Y_2$ interaction. Hence, $Y_1$ and $Y_2$ are initially uncorrelated, and the $XY_2$ interaction decreases the $XY_1$ correlations. For $\theta_{Y_1Y_2}\approx\pi/4$, there is now a tradeoff. The $Y_1Y_2$ interaction will decrease the $XY_1$ correlations, but it will also generate $Y_1Y_2$ correlations. If $\theta_{XY_2}$ is large, the $XY_2$ interaction can take advantage of the $Y_1Y_2$ correlations and result in greater $XY_1$ correlations.

For example, for $\theta_{XY_2}=0.8(\pi/2)$, (the bottom plots of Fig.~\ref{fig:finals1011}), the decrease in $XY_1$ correlations during the $Y_1Y_2$ interaction followed by the increase (or simply smaller decrease) during the $XY_2$ interaction may result in a smaller overall decrease than if there was no $Y_1Y_2$ interaction at all and only the $XY_2$ interaction. This point is a bit subtle and may require some contemplation. 

The overall result is that at larger values of $\theta_{XY}$, the peak of $G_{\infty}$ at fixed $\theta_{Y_2Y_1}$ may be at relatively high $\theta_{Y_1Y_2}$ values, as seen in Fig.~\ref{fig:finals1011}. For small $\theta_{XY_2}$ (top row of Fig.~\ref{fig:finals1011}), we only get the decrease in $XY_1$ correlations due to the $Y_1Y_2$ interaction with no increase due to the $XY_2$ interaction.

\subsection{Varying $\theta_{XY}=\theta_{XY_1}=\theta_{XY_2}$ and $\theta_{YY}=\theta_{Y_1Y_2}=\theta_{Y_2Y_1}$}
\begin{figure}
\centering
\includegraphics[scale=0.49]{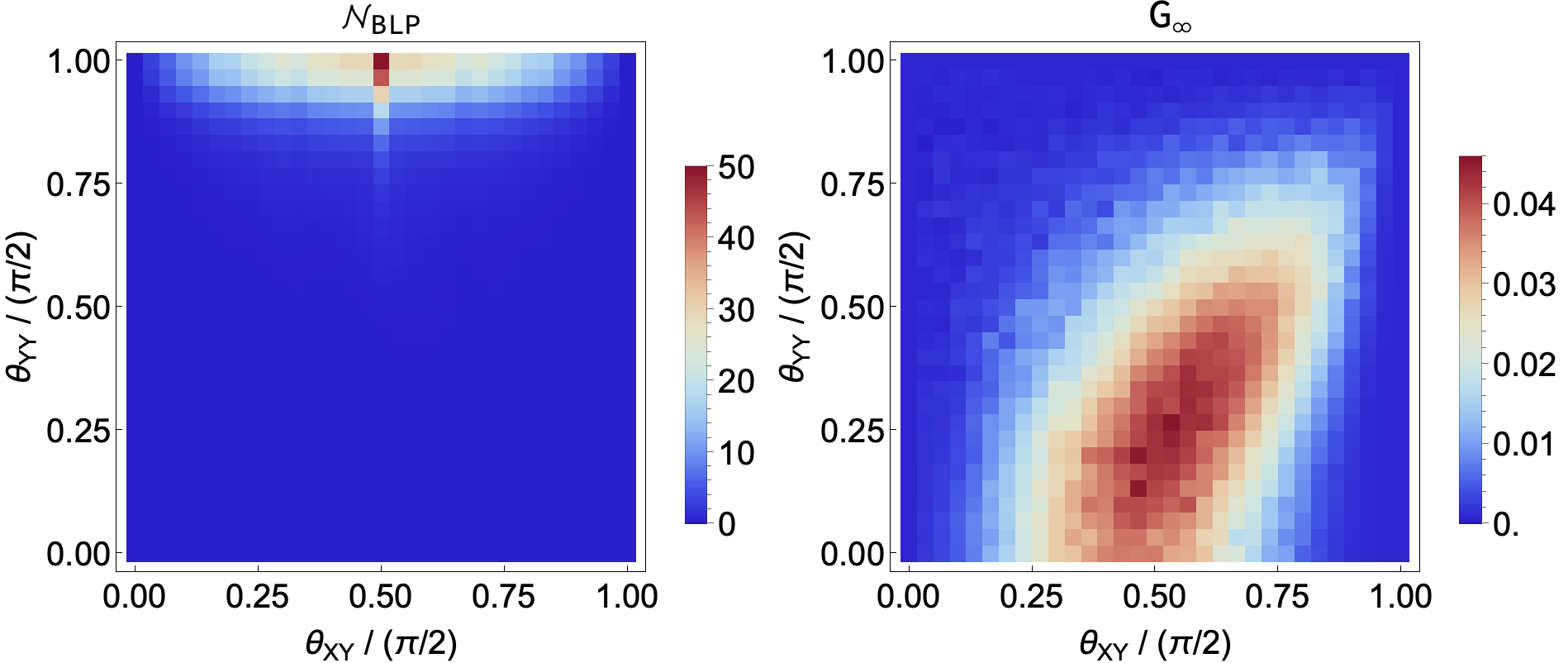}
\caption{$\mathcal{N}_{\mathrm{BLP}}$ and $G_{\infty}$ are shown for various values of $\theta_{XY}=\theta_{XY_1}=\theta_{XY_2}$ and $\theta_{YY}=\theta_{Y_1Y_2}=\theta_{Y_2Y_1}$, with $\rho_{X_0}=\ket{+}\bra{+}$ and $\rho_Y=\rho_{\mathrm{th}}(0.3)\otimes\ket{+}\bra{+}$. Spearman's $\rho$ for $G_{\infty}$ and $\mathcal{N}_{\mathrm{BLP}}$ is approximately $-0.31$ ($p<10^{-20}$).}
\label{fig:finals2}
\end{figure}

Here we look at the effects of varying both the $XY$ and $YY$ interaction parameters, always setting $\theta_{XY_1}=\theta_{XY_2}$ and $\theta_{Y_1Y_2}=\theta_{Y_2Y_1}$. The results for $\mathcal{N}_{\mathrm{BLP}}$ and $G_{\infty}$ are shown in Fig.~\ref{fig:finals2}. Note that some of the results we already discussed are embedded in these plots. For example, the verticals at $\theta_{XY}=0.2(\pi/2)$ and $\theta_{XY}=0.8(\pi/2)$ in Fig.~\ref{fig:finals2} correspond to the $\theta_{Y_1Y_2}=\theta_{Y_2Y_1}$ diagonal in the top and bottom plots of Fig.~\ref{fig:finals1011}, respectively. Also, the horizontal at $\theta_{YY}=0$ in Fig.~\ref{fig:finals2} corresponds to the $\theta_{XY_1}=\theta_{XY_2}$ diagonal of Fig.~\ref{fig:finals1}.

\subsubsection{\boldmath $G_\infty$}

The $G_{\infty}$ plot of Fig.~\ref{fig:finals2} more clearly demonstrates the subtle point that we just discussed. At low values of $\theta_{XY}$, the $YY$ interaction simply causes the $XY_1$ correlations to decrease, and the $XY_2$ interaction is too weak for the system to take advantage of the $Y_1Y_2$ correlations formed. Therefore, increasing $\theta_{YY}$ results in a monotonic decrease in $G_{\infty}$. However, at high values of $\theta_{XY}$, the system can take advantage of the $Y_1Y_2$ correlations, so that increasing $\theta_{YY}$ can result in higher values of $G_{\infty}$. However, eventually, the decrease due to the $Y_1Y_2$ interaction overpowers the increase due to the $XY_2$ interaction, and $G_{\infty}$ decreases.

On the other hand, considering the horizontals, i.e., fixed $\theta_{YY}$, since $\theta_{XY_1}=\theta_{XY_2}$, the competition between the two interactions at determining the steady state is irrelevant. Recall that $\theta_{XY_1}\approx\pi/4$ maximizes the $XY_1$ correlations formed during the $XY_1$ interaction, but at low values of $\theta_{YY}$, increasing $\theta_{XY_2}$ monotonically decreases the correlations. Therefore, we expect the peak to be close to but a bit less than $\theta_{XY}=\pi/4$. As mentioned, at higher values of $\theta_{YY}$, the system may be able to take advantage of the $Y_1Y_2$ correlations if $\theta_{XY}$ is high. All of this results in the observed right shifted elliptical contours. Understanding all of this requires careful consideration of what occurs at each interaction of a time step.

\subsubsection{\boldmath $\mathcal{N}_{\mathrm{BLP}}$}
\begin{figure}
\centering
\includegraphics[scale=0.49]{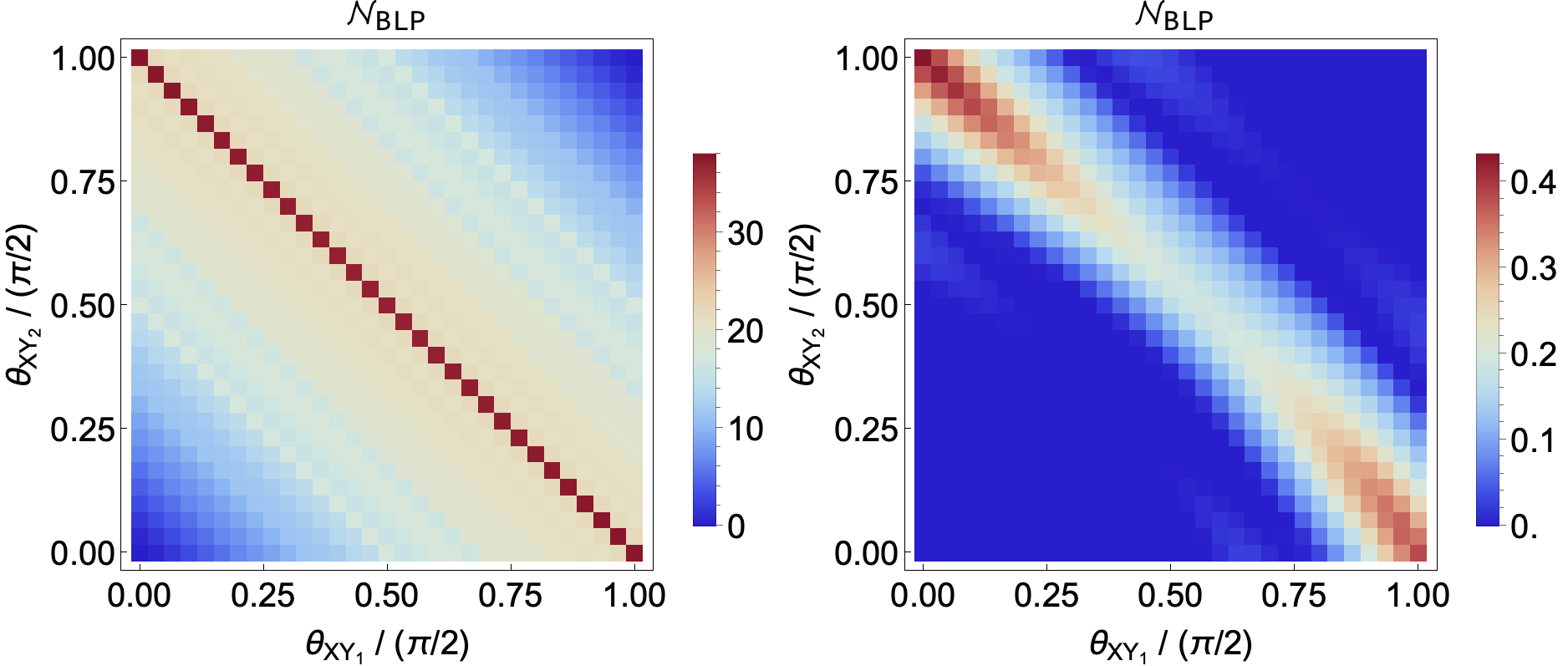}
\caption{$\mathcal{N}_{\mathrm{BLP}}$ is shown for various values of $\theta_{XY_1}$ and $\theta_{XY_2}$, with $\theta_{Y_1Y_2}=\theta_{Y_2Y_1}=0.95(\pi/2)$ (left-hand plot) and $\theta_{Y_1Y_2}=\theta_{Y_2Y_1}=0.5(\pi/2)$ (right-hand plot). We set $\rho_{X_0}=\ket{+}\bra{+}$ and $\rho_Y=\rho_{\mathrm{th}}(0.3)\otimes\ket{+}\bra{+}$.}
\label{fig:finals16}
\end{figure}

To understand the $\mathcal{N}_{\mathrm{BLP}}$ plot of Fig.~\ref{fig:finals2}, first consider Fig.~\ref{fig:finals16}, which shows $\mathcal{N}_{\mathrm{BLP}}$ for various values of $\theta_{XY_1}$ and $\theta_{XY_2}$, with $\theta_{Y_1Y_2}=\theta_{Y_2Y_1}=0.95(\pi/2)$ (left-hand plot) and $\theta_{Y_1Y_2}=\theta_{Y_2Y_1}=0.5(\pi/2)$ (right-hand plot). 

For $\theta_{Y_1Y_2}=\theta_{Y_2Y_1}=0.95(\pi/2)$, there is a very clear symmetry around the line $\theta_{XY_1}+\theta_{XY_2}=\pi/2$, with $\mathcal{N}_{\mathrm{BLP}}$ maximal on the line. The reason for this is as follows. As $\theta_{Y_1Y_2} \rightarrow\pi/2$, the $Y_1Y_2$ interaction swaps $Y_1$'s state with $Y_2$'s state. Therefore, the system is effectively only interacting with one qubit at each time step, with an interaction parameter given by $\theta_{XY_1}+\theta_{XY_2}$. When $\theta_{XY_1}+\theta_{XY_2}=\pi/2$ and $\theta_{Y_1Y_2},\theta_{Y_2Y_1} \rightarrow\pi/2$, the system effectively alternates between its initial state and the initial state of $Y_1$. Therefore, $\mathcal{N}_{\mathrm{BLP}}$ is given by the following diverging sum:

\begin{equation}
\mathcal{N}_{\mathrm{BLP}}
= 
\sum_{t=2,4,\ldots}
\Big[
D(\rho_{1,0}, \rho_{2,0})
\Big].
\end{equation}

For $\theta_{Y_1Y_2}=\theta_{Y_2Y_1}=0.95(\pi/2)$, the sum eventually converges, but very high degrees of non-Markovianity are reached as $\theta_{XY_1}+\theta_{XY_2}\rightarrow \pi/2$. The symmetry around $\theta_{XY_1}+\theta_{XY_2}=\pi/2$ is essentially due to the symmetry of $\sin(\theta)$ around $\pi/2$. Decreasing or increasing the sum from $\pi/2$ essentially pushes the overall interaction to $\mathbb{I}$ or $-\mathbb{I}$, and the phase is irrelevant.

It may be a bit strange that $\mathcal{N}_{\mathrm{BLP}}=0$ at $\theta_{XY_1}=\theta_{XY_2}=\pi/2$ for any $YY$ interaction parameters. However, in this case $Y_1$'s state is swapped to $X$ and then to $Y_2$ in the $XY_1$ and $XY_2$ interactions, respectively. Hence, whatever values $\theta_{Y_1Y_2}$ and $\theta_{Y_2Y_1}$ take, the $Y_2Y_1$ interaction will always be a partial swap between two identical states. Therefore, no information or correlations can be passed to the next ancilla, and the trace distance can never increase. 

Comparing with $\theta_{Y_1Y_2}=\theta_{Y_2Y_1}=0.95(\pi/2)$, the same symmetry is seen for  $\theta_{Y_1Y_2}=\theta_{Y_2Y_1}=0.5(\pi/2)$, albeit less pronounced (see Fig.~\ref{fig:finals16}). Also, the values of $\mathcal{N}_\mathrm{BLP}$ are roughly two orders of magnitude smaller. Also, along the line $\theta_{XY_1}+\theta_{XY_2}=\pi/2$, values further from the center are greater, i.e., $\mathcal{N}_{\mathrm{BLP}}$ at $\theta_{XY_1}=0$, $\theta_{XY_2}=\pi/2$ is greater than at $\theta_{XY_1}=\theta_{XY_2}=\pi/4$. This makes sense, because the weaker $YY$ interactions mean that both ancilla qubits are relevant, and $\theta_{XY_1}=\theta_{XY_2}=\pi/4$ does not equate to an effective full swap. 

Now, the $\mathcal{N}_{\mathrm{BLP}}$ plot of Fig.~\ref{fig:finals2} can be clearly understood. At fixed $\theta_{YY}$, the maximum at $\theta_{XY}=\pi/4$ is due to the total system--ancilla interaction approaching a full swap, particularly at high $\theta_{YY}$. At a fixed value of $\theta_{XY}$, intuitively, increasing $\theta_{YY}$ results in greater $\mathcal{N}_{\mathrm{BLP}}$, as more information and correlations can be passed along to the incoming ancilla.

\section{Summary and discussion}
\label{sec:conc}
Since the study of ISSs is still in its early stages, this thesis seeks to broaden the current understanding of the subject and shed light on memory effects. Often Markovianity is taken as an assumption, but with further advances in quantum technologies and experimental control, modeling quantum processes as accurately as possible is becoming increasingly important. 

We investigated how varying the interaction parameters affects $\mathcal{N}_{\mathrm{BLP}}$ and $G_{\infty}$. We motivated and then numerically showed that in the Markovian case, $G_{\infty}$ peaks at $\theta_{XY_1}$ slightly less than $\pi/4$ and $\theta_{XY_2}\approx\pi/4$. Then, we looked at the effects of varying $\theta_{Y_1Y_2}$ and $\theta_{Y_2Y_1}$. We saw that both need to be large to achieve large $\mathcal{N}_{\mathrm{BLP}}$. If one of them is small, this interaction will act as a bottleneck, preventing information and correlations from building up. 

Focusing on $G_{\infty}$, we saw that if one or both of $\theta_{Y_1Y_2}$ and $\theta_{Y_2Y_1}$ are large, the system effectively interacts with the same state over and over, and hence correlations will be minor in the steady state. We also saw that at high $\theta_{XY_1}=\theta_{XY_2}$, the decrease in $XY_1$ correlations during the $Y_1Y_2$ interaction can be offset by an increase during the $XY_2$ interaction. Lastly, because the system effectively interacts with 1-qubit ancillas at high $\theta_{YY}=\theta_{Y_1Y_2}=\theta_{Y_2Y_1}$ with an interaction parameter given by $\theta_{XY_1}+\theta_{XY_2}$, $\mathcal{N}_{\mathrm{BLP}}$ is maximized for $\theta_{XY_1}+\theta_{XY_2}\approx\pi/2$. This is also roughly seen at lower values of $\theta_{YY}=\theta_{Y_1Y_2}=\theta_{Y_2Y_1}$.

A statistically significant negative correlation between $G_{\infty}$ and $\mathcal{N}_{\mathrm{BLP}}$ was observed. This was anticipated based on the fact that if more information is flowing back to the system, less will be available for measurements. However, a caveat should be noted: the degree of non-Markovianity, $\mathcal{N}_{\mathrm{BLP}}$, is computed based on the approach to the steady state, while $G_{\infty}$ is computed in the steady state. Also, we remarked earlier on the fact that this correlation does not imply causation. It is entirely possible that in other models with various confounding factors, a statistically significant positive correlation could be seen. 

Several avenues for future work remain. In this thesis, non-Markovianity was introduced by allowing qubits initialized in the $Y_1$ state to interact with qubits initialized in the $Y_2$ state. However, other physically motivated mechanisms could also be considered. For example, one could envision two independent ancilla streams, in which $Y_1$ ($Y_2$) qubits interact with the next $Y_1$ ($Y_2$). Exploring how such alternative memory mechanisms affect $\mathcal{N}_{\mathrm{BLP}}$, $G_{\infty}$, and related conditional quantities would provide insight into the robustness and generality of the conclusions obtained in this work. However, this paper still provides a framework and guide that researchers can follow to understand how non-Markovian mechanisms affect their specific models.

Additionally,  Ref.~\cite{PRXQuantum.3.010303} introduced and discussed ISSs in the context of thermodynamics, with a focus on conditional entropy production. Therefore, a natural extension of this work would be to investigate how non-Markovian mechanisms influence the thermodynamic properties of informational steady states and their associated entropy production.

\bibliographystyle{ieeetr}
\bibliography{citations}

\end{document}